\documentclass[letter,twocolumn]{jpsj3}

\usepackage{graphicx}
\usepackage{color}

\begin{document}

\title{Pseudogap Behavior of the Nuclear Spin-lattice Relaxation Rate in FeSe Probed by $^{77}$Se-NMR}

\author{Anlu Shi$^1$\thanks{shi@scphys.kyoto-u.ac.jp}, Takeshi Arai$^1$, Shunsaku Kitagawa$^1$, Takayoshi Yamanaka$^1$, Kenji Ishida$^1$\thanks{kishida@scphys.kyoto-u.ac.jp}, Anna E. B\"{o}hmer$^2$, Christoph Meingast$^2$, Thomas Wolf$^2$, Michihiro Hirata$^3$, and Takahiko Sasaki$^3$}

\inst{$^1$Department of Physics, Graduate School of Science, Kyoto University, Kyoto 606-8502, Japan \\
$^2$Institute of Solid State Physics, Karlsruhe Institute of Technology, Karlsruhe D-76021, Germany \\
$^3$Institute for Materials Research, Tohoku University, Sendai 980-8577, Japan} 

\abst{We conducted $^{77}$Se-nuclear magnetic resonance studies of the iron-based superconductor FeSe in magnetic fields of 0.6 to 19 T to investigate the superconducting and normal-state properties. 
The nuclear spin-lattice relaxation rate divided by the temperature $(T_1T)^{-1}$ increases below the structural transition temperature $T_\mathrm{s}$ but starts to be suppressed below $T^*$, well above the superconducting transition temperature $T_\mathrm{c}(H)$, resulting in a broad maximum of $(T_1T)^{-1}$ at $T_\mathrm{p}(H)$.
This is similar to the pseudogap behavior in optimally doped cuprate superconductors.
Because $T^*$ and $T_\mathrm{p}(H)$ decrease in the same manner as $T_\mathrm{c}(H)$ with increasing $H$, the pseudogap behavior in FeSe is ascribed to superconducting fluctuations, which presumably originate from the theoretically predicted preformed pair above $T_\mathrm{c}(H)$.}


\maketitle

The discovery of superconductivity in iron pnictide\cite{KamiharaJACS2008} provided new research systems in which the unconventional superconductivity realized in strongly correlated electron compounds can be studied.
Among the Fe-based superconductors (FeSCs), FeSe, which exhibits superconductivity at $T_\mathrm{c} \sim 9$ K, has the simplest crystalline structure, which is called the 11 structure\cite{HsuPNAS2008}, but it exhibits several properties unlike those of other FeSCs.
A tetragonal-to-orthorhombic structural transition occurs at $T_\mathrm{s} \sim$ 90 K without long-range antiferromagnetic (AFM) ordering down to $T_\mathrm{c}$\cite{McQueenPRL2009}. 
This is in stark contrast with the other FeSCs, such as the 122 and 1111 systems, where static AFM ordering was observed at or slightly below $T_\mathrm{s}$\cite{IshidaJPSJ2009, JohnstonAdvPhys2010}. 
We reported that the $^{77}$Se nuclear magnetic resonance (NMR) spectrum splits below $T_\mathrm{s}$ and that the nuclear spin-lattice relaxation rate divided by $T$, $(T_1T)^{-1}$, of $^{77}$Se is enhanced below $T_\mathrm{s}$\cite{BoehmerPRL2015}. 
These results suggest that orbital ordering induces the electronic nematic state below $T_\mathrm{s}$ and also triggers the development of AFM fluctuations with stripe correlations breaking the $C_4$ symmetry in FeSe\cite{BoehmerPRL2015, BaekNatMatt2014}. 
This scenario is consistent with the results of angle-resolved photoemission spectroscopy\cite{NakayamaPRL2014, WatsonPRB2015}, but it seems to be inconsistent with the AFM-fluctuation-driven nematic scenario\cite{FernandesNatPhys2014}.
In any case, FeSe is an ideal system for studying the origin of the electronic nematic state in FeSCs.

It was recently suggested that the superconductivity in FeSe may be in the Bardeen--Cooper--Schrieffer (BCS)-Bose--Einstein condensate (BEC) crossover regime, as an extremely small Fermi energy comparable to the superconducting (SC) condensation energy was revealed by several measurements\cite{KasaharaPNAS2014, OkazakiSciRep2014}. 
In a schematic phase diagram of the attractive Hubbard model, the interaction between two fermions (e.g., electrons in solids) is in the weak-coupling regime in the BCS state and in the strong-coupling regime in the BEC state.
The characteristic parameter is the ratio of the SC gap $\Delta$ to the Fermi energy $\varepsilon_\mathrm{F}$, and $\Delta/\varepsilon_\mathrm{F}\sim1$ in the crossover regime\cite{RanderiaARCMP2014, ChenPhysRep2005}. 
The most intriguing property in this crossover regime is the formation of a theoretically predicted preformed pairing state of fermions. 
Although preformed pairing in YBa$_2$Cu$_3$O$_7$\cite{LortzPRL2003} has been suggested, the nature of the preformed pairing is far from clear.
A recent study of FeSe by magnetic torque measurement provides evidence of the existence of preformed pairs below $T^{*}\sim$ 20 K, which is shown by a large enhancement in the SC fluctuations ascribed to the preformed pairs\cite{KasaharaNatComm2016}. Therefore, FeSe is one of the best superconductors for studying the properties of preformed pairs.

We studied FeSe using $^{77}$Se NMR measurements in various magnetic fields to investigate the SC properties and the interplay between the superconductivity and AFM fluctuations microscopically.
In this Letter, we focus on the magnetic field ($H$) dependence of $T_1^{-1}$ near $T_\mathrm{c}$, as the unconventional behavior anticipated in the strong-coupling-limit regime would be observed in this temperature region.
We found that suppression of AFM fluctuations begins below $T^*$, resulting in a broad maximum of $(T_1T)^{-1}$ at $T_\mathrm{p}(H)$ above the SC transition temperature $T_\mathrm{c}(H)$.  
This behavior is reminiscent of the pseudogap (PG) behavior observed in the optimally doped YBa$_2$Cu$_3$O$_7$.
From the $H$ dependence of $T_\mathrm{p}$ and $T_\mathrm{c}$, we claim that the PG behavior in FeSe is attributable to the SC fluctuation effects, which suppress the density of states (DOS) by forming preformed pairs above $T_\mathrm{c}$. We develop the phase diagram of the SC fluctuation effect on the basis of some anomalies and kinks in the temperature and magnetic field dependence of ($T_1T$)$^{-1}$.

The single crystalline sample used in the measurements was grown by a low-temperature vapor transport method\cite{BoehmerPRB2013}. The 
$T_\mathrm{c}(H)$ value of the sample was determined by measuring a diamagnetic shielding signal detected by an NMR coil, which enabled us to compare the NMR anomalies with the SC transition directly. 
$T_\mathrm{c}(0)$ is $\sim 9$ K, which is consistent with previous studies\cite{BoehmerPRB2013}. 
Because the $^{77}$Se nucleus has $I = 1/2$, the nuclear quadrupolar effect is absent; thus, an external field is needed for the NMR experiments, and it is a good probe for detecting magnetic anomalies microscopically. 
Magnetic fields were applied along the \textit{c} axis in the present measurements, as the superconductivity is effectively suppressed in this direction, and the $H$-induced normal state can be observed in a wide range of $T$.
The $H$-induced normal-state properties were investigated by NMR measurements under high magnetic fields ($\mu_0H \sim 19$ T) exceeding the upper critical magnetic field $\mu_{0}H_\mathrm{c2}(0)$ at the High Field Laboratory for Superconducting Materials, Institute for Materials Research, Tohoku University.
The details of the experiments are summarized in the Supplementary Materials\cite{SM}.

\begin{figure}
\includegraphics[width=\linewidth]{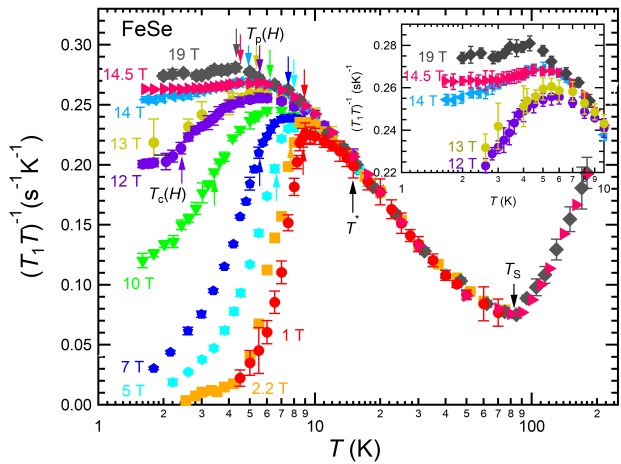}
  \caption{(Color online) Temperature dependence of $(T_{1}T)^{-1}$ in various magnetic fields. The structural transition temperature $T_\mathrm{s}$ is indicated by a downward black arrow. $T_\mathrm{c}(H)$ and $T_\mathrm{p}(H)$ are indicated by upward and downward arrows, respectively, in various colors (see the text). (Inset) Expanded view of $(T_1T)^{-1}$ at 12, 13, 14, 14.5, and 19 T below $T=$ 10 K.}
  \label{figure1}
\end{figure}

The $^{77}$Se nuclear spin-lattice relaxation rate $T_{1}^{-1}$ is obtained by fitting the recovery curve of the nuclear magnetization with the formula $[M(\infty)-M(t)]/M(\infty) \propto \exp(-t / T_{1})$ for an $I=1/2$ nucleus.
Here, $M(t)$ is the nuclear magnetization at a time $t$ after a saturation pulse. A single component of $T_{1}^{-1}$ is related to the dynamical susceptibility and is expressed by the formula
\begin{equation}
  \frac{1}{T_{1}}=\frac{2\gamma_{n}^{2}k_\mathrm{B}T}{(\gamma_{e}\hbar)^{2}}\sum_{\textbf{\textit{q}}}|A_{\textbf{\textit{q}}}|^{2}\frac{\chi_{\perp}^{''}(\textbf{\textit{q}}, \omega_{0})}{\omega_{0}},
\end{equation}
where $\gamma_n$ ($\gamma_e$) is the gyromagnetic ratio of the nuclear spins (electronic spins), $A_{\textbf{\textit{q}}}$ is the hyperfine coupling tensor, $\chi_{\perp}^{''}(\textbf{\textit{q}}, \omega_{0})$ is the imaginary part of the dynamical susceptibility in the direction perpendicular to the quantization axis, and $\omega_{0} = \gamma_{n}H$ is the resonance frequency under a magnetic field $\mu_{0}H$. 
Because $(T_{1}T)^{-1}$ is proportional to the $q$-summed $\chi_{\perp}^{''}(\textbf{\textit{q}}, \omega_{0})$, and $(T_{1}T)^{-1} \propto N(E_\mathrm{F})^2$ values of conventional metals, where $N(E_\mathrm{F})$ is the DOS at the Fermi level, the value of $(T_{1}T)^{-1}$ is a good quantity for determining the properties of local low-energy (MHz--$\mu$eV range) magnetic fluctuations and the SC gap.

The temperature dependence of $(T_{1}T)^{-1}$ under different magnetic fields is shown in Fig.\ \ref{figure1}. 
$(T_{1}T)^{-1}$ decreases from room temperature to $T_\mathrm{s}$. 
As reported in previous papers\cite{BaekNatMatt2014, BoehmerPRL2015}, low-energy AFM fluctuations develop gradually below $T_\mathrm{s}$, as seen in Fig.\ \ref{figure1}.   
The onset temperature below which AFM fluctuations develop is unchanged up to 19 T, indicating that $T_\mathrm{s}$ is independent of $H$ in this magnetic field range.
The development of AFM fluctuations is independent of $H$ down to $T=15$ K, below which $(T_1T)^{-1}$ is observed to be $H$-dependent.
As $T$ approaches $T_\mathrm{c}(H)$, $(T_{1}T)^{-1}$ shows a broad maximum at $T_\mathrm{p}(H)$, which will be discussed later.

\begin{figure}
\includegraphics[width=\linewidth]{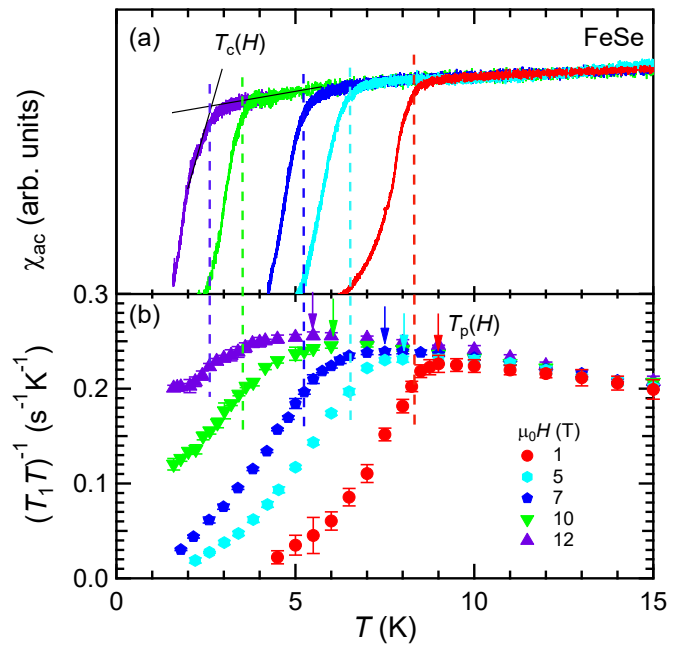}
\caption{(Color online) Temperature dependence of (a) diamagnetic shielding signals and (b) $(T_{1}T)^{-1}$ in various magnetic fields below 15 K. $T_\mathrm{c}(H)$ is determined by linear fitting of the diamagnetic signals around the turning point and is indicated by dashed lines. $T_\mathrm{p}$ is determined by the temperature of maximum $(T_{1}T)^{-1}$ and is indicated by downward arrows. }
  \label{figure2}
\end{figure}

Clear evidence of an SC precursor was observed in the magnetic fields.
Figure \ref{figure2} compares the diamagnetic shielding signal and $(T_1T)^{-1}$ below $T=$ 15 K.
With decreasing temperature, $(T_1T)^{-1}$ shows a broad maximum at $T_\mathrm{p}$.
$T_\mathrm{p}$ is slightly higher than $T_\mathrm{c}$($H$) and is $H$-dependent.
The temperature region between $T_\mathrm{p}$ and $T_\mathrm{c}$ becomes wider with increasing $H$, indicating that the SC fluctuation effect becomes significant with increasing $H$.
The broad maximum of $(T_1T)^{-1}$ just above $T_\mathrm{c}$ is similar to the PG behavior observed in YBa$_2$Cu$_3$O$_7$\cite{WarrenPRL1989, Yasuoka1989}.
Because $T_\mathrm{p}$ and $T_\mathrm{c}$ are suppressed by $H$ in the same manner, as shown later, it is suggested that the PG behavior in FeSe can be ascribed to the SC properties.

\begin{figure}
\includegraphics[width=\linewidth]{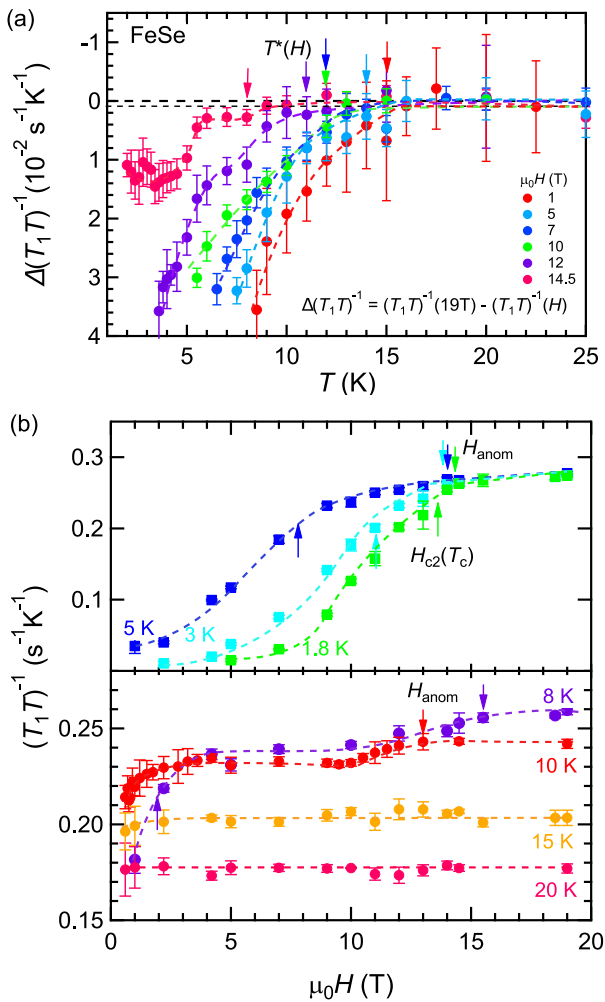}
\caption{(Color online) (a) $T$ dependence of difference between $(T_1T)^{-1}$ at $\mu_0H=$ 19 T and $(T_1T)^{-1}$ in other magnetic fields. $T^*(H)$ is shown by downward arrows. (b) $H$ dependence of $(T_1T)^{-1}$ at various temperatures. $H_\mathrm{c2}(T_\mathrm{c})$ and $H_\mathrm{anom}$ are indicated by upward and downward arrows, respectively. }
\label{figure3}
\end{figure}

To determine the temperature below which the $H$ dependence of $(T_1T)^{-1}$ emerges, we plotted $\Delta(T_1T)^{-1} = (T_1T)^{-1}$(19 $\mathrm{T}) - (T_1T)^{-1}(H)$ in Fig.\ \ref{figure3}(a), as $\mu_0H=$ 19 T is the highest field at which the SC transition is suppressed. 
Note that $T^*$ should be zero at $\mu_0H=$ 19 T according to the definition of $T^*$, which indicates the ambiguity of the behavior of $T^*$ in higher-field regions.
For convenience, we determine the $T^*(H)$ value below which $\Delta(T_1T)^{-1}$ exceeds $1 \times 10^{-3}(\mathrm{sK})^{-1}$, as indicated in Fig.\ \ref{figure3}(a).
We can safely say that $T^*(H)$ is the onset temperature of the PG state, at which suppression of AFM fluctuations begins.

We also performed field scan measurements of $(T_1T)^{-1}$ at various temperatures.
Figure \ref{figure3}(b) shows the $H$ dependence of $(T_1T)^{-1}$.
At $T=$ 10 K, near $T_\mathrm{c}(0)$, $(T_1T)^{-1}$ is definitely suppressed toward lower $H$ below $\mu_0H=$ 3 T. 
This behavior is in strong contrast to that at $T=$ 20 K, indicating the presence of SC fluctuations.
In the high-$H$ region, $(T_1T)^{-1}(H)$ at 10 K has another kink at $H_\mathrm{anom}\sim$ 13 T.
A similar anomaly can also be observed below 10 K. 
$(T_1T)^{-1}(H)$ increases monotonically toward $H_\mathrm{anom}$, passing through $H_\mathrm{c2}(T_\mathrm{c})$, and saturates above $H_\mathrm{anom}$, as shown in Fig.\ \ref{figure3}(b).
Note that in the high-$H$ region above $H_\mathrm{anom}$, the large reduction of $(T_{1}T)^{-1}$ below $T_\mathrm{p}$ is not observed, but $(T_{1}T)^{-1}$ is almost constant at low $T$, as shown in the inset of Fig.\ \ref{figure1}.
This suggests that the SC fluctuation effect is weak in the high-$H$ region above $H_\mathrm{anom}$.

\begin{figure}[t]
\includegraphics[width=\linewidth]{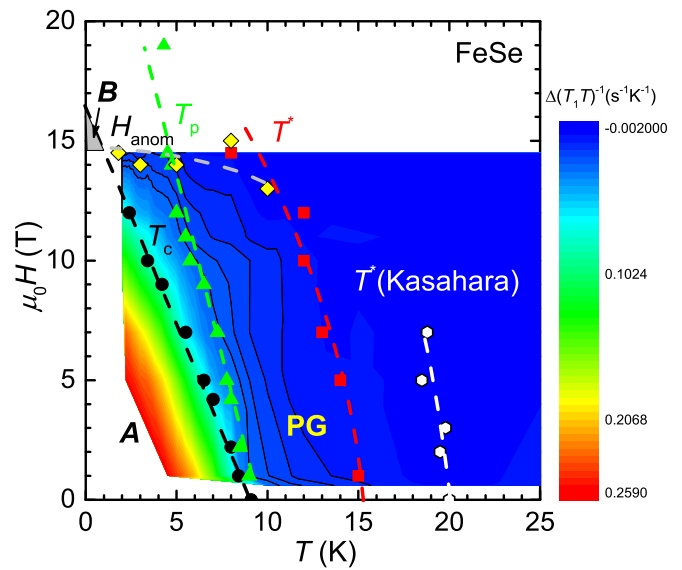}
 \caption{(Color online) $H-T$ phase diagram of FeSe for $H \parallel c$. Black circles and green triangles represent $T_\mathrm{c}(H)$ and $T_\mathrm{p}(H)$, respectively. $T^*(H)$ and $H_\mathrm{anom}$ are plotted as red squares and yellow filled diamonds, respectively. $T^*$ values from Kasahara, determined from thermodynamics and magnetic torque measurements, are also plotted as white hexagons\cite{KasaharaNatComm2016}. The $B$ phase\cite{KasaharaPNAS2014} is also shown in the diagram as a filled gray triangular region. The rest of the SC phase is identified as the $A$ phase.}
\label{figure4}
\end{figure}

On the basis of the temperature and field dependence, $T_\mathrm{c}(H)$, $T_\mathrm{p}(H)$, $T^*(H)$, and $H_\mathrm{anom}$ are plotted on the $H-T$ phase diagram for $H$ $\|$ $c$ in FeSe, as shown in Fig.\ \ref{figure4}.
The values of $\Delta(T_1T)^{-1}$ in Fig.\ \ref{figure3}(a) are also illustrated as a contour plot in the phase diagram.
The $H$-induced unconventional SC $B$ phase reported by Kasahara\cite{KasaharaPNAS2014} and Watashige\cite{WatashigeJPSJ2017} is also shown in the same figure.
It seems that $H_\mathrm{anom}$ is linked to the $A-B$ phase transition line in the SC state.
On the other hand, the $H$ dependence of $T_\mathrm{p}(H)$ and $T^*$ is similar to that of $T_\mathrm{c}(H)$ up to $\mu_{0}H=$ 10 T, but it seems that $T^*$ shows different behavior from $T_\mathrm{p}$ in the higher-field region above 12 T. This difference might originate from the definition of $T^*$, and $(T_1T)^{-1}$ measurement above 19 T is needed to determine the intrinsic behavior of $T^*$ in the high-field region.

Now, we discuss the origin of the suppression of $(T_1T)^{-1}$ below $T^*(H)$ in FeSe.
Note that the SC diamagnetic fluctuations due to the Aslamazov--Larkin process do not play a role in the $T_1^{-1}$ measurements, as they alter the magnetic field mainly along the axis parallel to the applied field, and only transverse fields contribute to the relaxation of the $c$-axis component of the nuclear spin\cite{RanderiaPRB1994}.   
Magnetic fields generally tend to enhance pairing fluctuations near $T_\mathrm{c}(H)$ as a result of Landau quantization of the orbital motion of the pairs, but at the same time, they reduce the pairing fluctuations at constant temperature with increasing $H$, as the transition temperature is suppressed by $H$.
Application of a magnetic field at constant temperature just above $T_\mathrm{c}$ has very different effects on the pairing fluctuation contributions to $(T_1T)^{-1}$ depending on the pairing symmetry.
Theoretical studies showed that, for ordinary $s$-wave pairing, $(T_1T)^{-1}$ decreases with increasing $H$ owing to the anomalous Maki--Thompson effect, whereas in $d$-wave pairing, suppression of the pairing fluctuation effect leads to an increase in $(T_1T)^{-1}$ with $H$, as the pairing fluctuation effect originating from the preformed pairing reduces the normal-state DOS above $T_\mathrm{c}$\cite{EschrigPRB1999}.

The SC fluctuation effect can be estimated from $\Delta (T_1T)^{-1}$ in Fig.\ \ref{figure3}(a).
The value of $\Delta(T_1T)^{-1}/(T_1T)^{-1}$ at 10 K just above $T_\mathrm{c}$ is evaluated as $-0.086$ at $\mu_0H = $ 1 T, which is comparable to the experimental value for YBa$_2$Cu$_3$O$_7$ at 95 K and $\mu_0 H =$ 10 T [$\Delta(T_1T)^{-1}/(T_1T)^{-1} \sim -0.12$]\cite{MitrovicPRL1999}. 
This indicates that the SC fluctuation effect in FeSe has the same sign as that observed in $d$-wave pairing and the same magnitude as in YBa$_2$Cu$_3$O$_7$, although the temperature scale is one order of magnitude different in the two compounds, and FeSe is considered to have s$^{\pm}$ pairing symmetry.   
Here it is interesting to note that whereas YBa$_2$Cu$_3$O$_7$ also exhibits very strong 3D XY-type SC fluctuations in zero magnetic field\cite{PaslerPRL1998}, the specific heat anomaly in FeSe looks surprisingly mean-field-like\cite{LinPRB2011, WangPSSB2017, YangPRB2017}.

As mentioned in the introduction, the SC pairing fluctuations related to the preformed pairs become remarkable in the BCS-BEC crossover region, where the pairing interaction is so strong that it is comparable to the Fermi energy $\varepsilon_\mathrm{F}$.
In FeSe, the characteristic temperature of the magnetic fluctuations is regarded as $T_\mathrm{s}\sim$ 90 K, which is of the same order as the Fermi energy revealed by scanning tunneling microscopy measurements\cite{KasaharaPNAS2014}, and thus $\Delta/ \varepsilon_\mathrm{F} \sim 0.1$. 
Note that $\Delta/ \varepsilon_\mathrm{F} \sim 0.1$ is also comparable to the value in YBa$_2$Cu$_3$O$_7$ ($\Delta/ \varepsilon_\mathrm{F} \sim 0.05$), and the observation of the SC fluctuation effect in FeSe is understood with respect to the $\Delta/ \varepsilon_\mathrm{F}$ parameter.  
Because the PG formation associated with the preformed pairs, which leads to the suppression of low-energy single-particle excitation, has been theoretically predicted in this crossover region, it is reasonable to consider that the suppression of the AFM fluctuations below $T^*(H)$ and the decrease in $(T_1T)^{-1}$ below $T_\mathrm{p}(H)$ originate from preformed-pair formation without long-range phase coherence.

Although $T^*$ seems to connect to $H_\mathrm{anom}$, the $T^*$ line should cover the SC state because $T^*$ is regarded as the onset temperature of pairing formation.
$H_\mathrm{anom}$ seems to be related to the reported $A-B$ SC phase transition, as shown in Fig.\ \ref{figure4}.
We suggest that the SC characteristics of the $A$ and $B$ phases are different with respect to the presence of the SC fluctuations.
Recently, Watashige \textit{et al.} measured the thermal transport properties of the $B$ phase and suggested that the SC $B$ phase may be the unconventional Fulde--Ferrell--Larkin--Ovchinnikov (FFLO) phase\cite{WatashigeJPSJ2017}.
The weak SC fluctuations may be plausible for the FFLO state, as the superconductivity is almost destroyed in this state. 
To unveil the SC properties of the $B$ phase, NMR experiments on the $B$ phase are needed.

In conclusion, our $^\mathrm{77}$Se NMR measurement, which is the first microscopic study of SC fluctuations in this material, clarified the presence of the PG state and SC fluctuations in FeSe.
On the basis of the $H$ and $T$ dependence of $(T_1T)^{-1}$, we mapped an $H-T$ phase diagram of FeSe with respect to the SC fluctuation effect.
We also found an anomaly at high $H$ above $T_\mathrm{c}$, which might be related to the $A-B$ SC phase transition.

\begin{acknowledgment}


We thank Y. Matsuda, T. Shibauchi, S. Kasahara, K. Adachi, R. Ikeda, and Y. Yanase for valuable discussions. 
This work was partially supported by the Kyoto University LTM center and JSPS KAKENHI (Grant Numbers JP15H05882, JP15H05884, JP15K21732, JP15H05745, and JP17K14339.)
Part of this work was performed at the High Field Laboratory for Superconducting Materials, Institute for Materials Research, Tohoku University.

\end{acknowledgment}

\end{document}